# Navigating the Research Landscape of Decentralized Autonomous Organizations: A Research Note and Agenda

**Christian Ziegler**
Technical University of Munich
Arcisstraße 21, 80333 Munich
christian.ziegler@tum.de

**Quinn DuPont**
Schulich School of Business, York University
4700 Keele Street, Toronto
qdupont@yorku.ca

*This note and agenda serve as a cause for thought for scholars interested in researching Decentralized Autonomous Organizations (DAOs), addressing both the opportunities and challenges posed by this phenomenon. It covers key aspects of data retrieval, data selection criteria, issues in data reliability and validity such as governance token pricing complexities, discrepancy in treasuries, Mainnet and Testnet data, understanding the variety of DAO types and proposal categories, airdrops affecting governance, and the Sybil problem. The agenda aims to equip scholars with the essential knowledge required to conduct nuanced and rigorous academic studies on DAOs by illuminating these various aspects and proposing directions for future research.*

# 1   The Evolution of DAO Research

Decentralized Autonomous Organizations (DAOs) have emerged from blockchain-based infrastructure. In recent years, these new "online communities" have gained the focus of researchers trying to understand the complexities behind them. DAOs have rapidly evolved and display immense diversity today, making it challenging for scientists to define and analyze them. For example, there are DAOs that support critical infrastructure (MakerDAO is responsible for maintaining a complex, algorithmic system to dynamically adjust the token price of DAI, one of crypto's most important digital assets) and DAOs that mostly just party (e.g., ApeCoin DAO is responsible for maintaining the Bored Ape Club's NFTs and their "in real life" events).

However, the definitions of DAOs have also evolved with them. In 2013, Buterin (2013) introduced the idea that companies could evolve into decentralized autonomous corporations (DACs). The next year, he further elaborated on this idea, describing DAOs as entities that live on the internet and exist autonomously (Buterin, 2014). Soon after, Wright and Filippi (2015) described DAOs as software-based entities that can own and trade resources while mimicking traditional corporate governance. By this point, a few real DAOs were beginning to emerge (e.g., Dash), but famously it was not until the 2016 launch of "The DAO" (which was immediately attacked and exploited), when scholars began to pay attention to the risks and potentials of DAOs. By 2017, DAOs began to be recognized for their potential to operate autonomously, emphasizing decentralized decision-making (Dhillon et al., 2017). One year later, the definitions shifted to practical implications, emphasizing real-world business operations and service offerings (Beck et al., 2018; Hsieh et al., 2018). For instance, S. Wang et al. (2019) emphasize the lack of any central authority or management hierarchy for DAOs, however, such claims have never been truly fulfilled and most DAOs remain highly centralized and hierarchical. By 2020, DAOs were recognized for having a community that organizes itself on decentralized infrastructure (El Faqir et al., 2020). Still evolving, in 2021, some researchers shifted their focus to the ways that DAOs have become complex systems that mediate interactions between humans and blockchains and are striving to become entirely autonomous systems with storage and transaction of value, voting functions, autonomous execution of code, and decentralized security infrastructure (Hassan & Filippi, 2021; Rikken et al., 2021; Schillig, 2021).

At the core of every DAO, there are Treasury, Governance, and Community components (Ziegler & Welpe, 2022). In the Treasury, common funds are stored as digital assets, or 'tokens.' For Governance, proposals for change are created and voted on by the community, which must safeguard the treasury's funds, simultaneously protecting against asset exploitation while investing in its future. Communities are networked assemblages of people that typically have low barriers to entry (token ownership is often sufficient to participate). As governance proposals are the data traces that lie between the treasury and the governing community, we focus on them here.

While scholars have many tools to investigate new phenomena like DAOs, robust research requires a deep theoretical understanding and careful data analysis. We aim to highlight common and uncommon pitfalls researchers should be aware of while conducting quantitative analyses on DAOs and detail current research gaps and emerging research agendas. Existing methodologies are discussed and new experimental ones suitable for research on DAOs are proposed.

In the next section, we present recent work on DAOs, highlighting their focal points, methods, empirical findings and shortcomings.

## 2 Recent Research on DAOs

The first discussion of a practical DAO starts with the whitepaper of Jentzsch (2016) and presents implementation details for "The DAO." Following its brief existence, scholars investigated nascent aspects of DAOs, specifically examining community decision making and proposing preventive measures for future mishaps, as DuPont (2017) and Andryukhin (2018) discussed. Meanwhile, studies by Kondova and Barba (2019), S. Wang et al. (2019), Kaal (2021), Chao et al. (2022), and Qin et al. (2023) delve into the functionalities of DAOs or compare them with existing organizational structures and explore their capabilities. Baninemeh et al. (2023) outline the design principles of DAOs, while Faqir-Rhazoui et al. (2021) and El Faqir et al. (2020) investigate the platforms on which these DAOs operate, alongside analyzing the distribution of their traffic. From a governance perspective, Marko and Kostal (2022) contemplate avenues for its enhancement within DAOs. Sharma et al. (2023) offer analytics of DAOs, explicitly focusing on governance aspects and discussions, while Chohan (2017) and Han et al. (2023) focus on governance issues. Zhao et al. (2022) found that strategic decisions positively impact platform operational performance under some circumstances in DAOs. Ziegler and Zehra (2023) studied practitioners' rankings of DAOs. Finally, the work of Ziegler and Welpe (2022) differentiates between various DAO types, extending the taxonomy work of Wright (2021).

Currently, research on DAOs is mainly qualitative or based on literature reviews (Andryukhin, 2018; Baninemeh et al., 2023; Chao et al., 2022; DuPont, 2017; Kaal, 2021; Kondova & Barba, 2019; Marko & Kostal, 2022; Qin et al., 2023; Sharma et al., 2023; S. Wang et al., 2019; Ziegler & Zehra, 2023) while only some studies are quantitative (El Faqir et al., 2020; Faqir-Rhazoui et al., 2021; Feichtinger et al., 2023; Fritsch et al., 2022; J. R. Jensen et al., 2021; Liu, 2023; Ziegler & Welpe, 2022).

Due to the sheer variety of DAOs (Ziegler & Welpe, 2022), qualitative studies and literature reviews should be looked at with skepticism when they make claims about DAOs as a whole, instead of as a diverse assemblage of self-constituting commons which sometimes have shared values, goals, and infrastructures. For instance, while the centrality of MakerDAO is clearly evident by the attention it receives in scholarly literature, it is far from representative of the diversity of DAOs. Moreover, most quantitative studies have significant flaws, such as low sample size (Ziegler & Welpe, 2022) and the measurement of early network effects and other sociologically unstable phenomena (El Faqir et al., 2020; Faqir-Rhazoui et al., 2021). The study of Zhao et al. (2022), for example, only targeted MakerDAO[1] and used one year of data in an effort to capture a highly dynamic and evolving community. Other, more recent quantitative studies on voting in DAOs use only a tiny fraction of available data. For example, Feichtinger et al. (2023) investigated the voting of 621 proposals in 21 on-chain DAOs; in 17 of these, they found that 10 participants have the majority of the voting power. Fritsch et al. (2022) investigated only 3 DAOs and 94 proposals. While this empirical research offers valuable insights to nascent phenomena, we must guard against drawing false conclusions, such as when a DAO can or should be considered autonomous or decentralized.

More broadly, scholars have also designed research agendas specifically for DAOs and proposed research into socio-material practices, interactions, human-machine agency, and institutional change (Santana & Albareda, 2022). For instance, Beck et al. (2018) propose a research agenda for governance in the blockchain economy with the dimensions of decision rights, accountability, and incentives.

## 3 Our Contribution

To contextualize the challenges and opportunities facing DAOs, we conduct an integrated literature review. This is a method "that reviews, critiques, and synthesizes representative literature on a topic in an integrated way such that new frameworks and perspectives on the topic are generated." (Torraco, 2005). We perform a two-step approach; first, we gather and critically analyze existing work on DAOs and extract definitions, methodologies, research gaps, and results. During this stage of the review, we gathered 61 articles on DAOs using Google Scholar[2] using the keywords *DAO, Decentralized Autonomous Organization, Blockchain governance, on-chain governance, algorithmic governance, and decentralized governance* and then performed a backward and forward reference search. Recognizing the nascent field and lack of established, high quality scholarly publication venues, we explicitly include pre-prints and in-progress manuscripts from arXiv[3] and SSRN,[4] but we also recognize that our literature search is not exhaustive.

In a second step, we interrogate existing DAO studies by comparing and critiquing data providers, research methods, and the robustness of conclusions. During this step, we experimentally evaluated numerous datasets and perform statistical measurements to assess the practical and theoretical challenges that remain.

### 3.1 Methodological Framework and Meta-Analysis

---

[1] https://makerdao.com/en/
[2] https://scholar.google.com/
[3] https://arxiv.org/
[4] https://www.ssrn.com/

In this section, we discuss our integrated literature review findings, focusing initially on the challenges of quantitative DAO analysis. We then delve into specific methods for handling DAO data, including data retrieval, selection, and data problem-solving strategies. Finally, we unveil our map for future research agendas.

## 3.2 Measuring Successful DAOs

There are few published studies that measure, analyze, or predict the performance and success of DAOs other than Zhao et al. (2022), which looked at a dimension of operational performance only. Predicting the performance of organizations has shown to be challenging (Bonaventura et al., 2020; Bose & Pal, 2006; Johnson & Soenen, 2003) and thus we expect it to be very hard to do for DAOs, too. Aside from highlighting research gaps (Santana & Albareda, 2022), scholars have not discussed core questions about this kind of research, especially regarding what 'good' performance for DAOs means. We recognize that while investment-focused DAOs may follow the goal of maximizing their returns (typically measured by token price), On-Chain Product and Service DAOs might also try to maximize their token price, Off-Chain Product and Service DAOs might prioritize growth and user retention, and Community DAOs focus on increasing social capital by onboarding, connecting, and educating as many members as possible (Ziegler & Welpe, 2022). Thus, the proximate goal of each DAO remains unclear; given the fluidity of control and rapid development, the performance and success metrics differ from DAO to DAO and cannot be uniformly applied across the landscape. In turn, empirical limits on the available datasets frustrate statistically significant analyses.

Echoing the concerns of DAO community members, most DAO researchers have focused on the key issue of decentralization. Decentralization is important for DAOs because blockchain infrastructures require the effective decentralization of block-processing to avoid security issues, like 51% attacks. However, researchers have failed to explain how these security concerns are manifested in community concerns, resulting in a scholarly echo-chamber that attempts to measure decentralization without an appropriate grounding in organizational theory. Indeed, many scholars have performed roughly the same measurement under different names:

- Braun et al. (2021) use the Herfindahl-Hirschman Index (HHI) $HHI=\sum_{i=1}^{N}s_i^2$

- Sharma et al. (2023) and Sun et al. (2023) measure Gini coefficients $G=\sum_{i=1}^{N}\sum_{j=1}^{N}|s_i-s_j| / 2N^2\mu$

- Liu (2023) study $n$-top players $\sum_{i=1}^{n}s_i$

- Goldberg and Schär (2023) conduct a 'whale' ($n=1$ top player) analysis.

These are formally similar measurements that reveal differential relationships. Specifically, HHI and the Gini coefficient both mathematically encapsulate the entire distribution of market shares (with HHI emphasizing the squared influence of each firm's size and Gini focusing on relative differences), the *n*-top player (and whale) analysis straightforwardly sums the largest shares, offering a more direct, though less nuanced, view of market concentration. We performed similar analyses on our dataset for comparison (see Figures 1, 2, and 3).

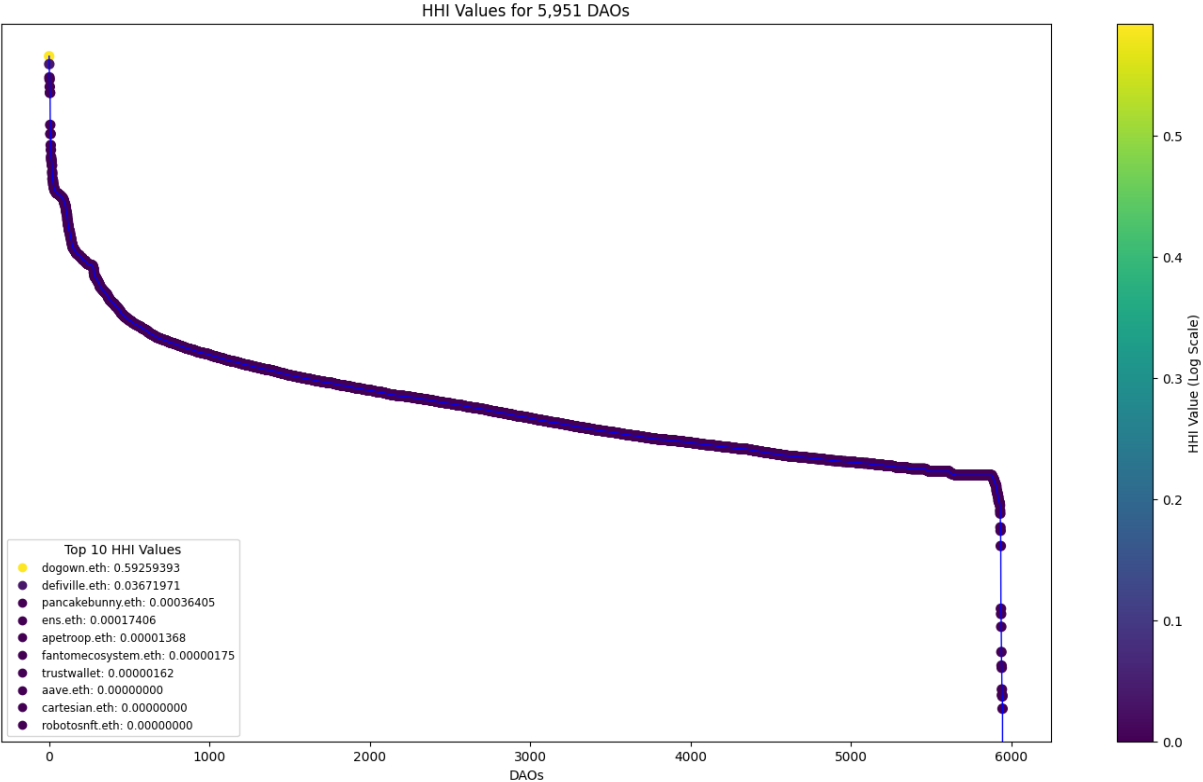

*Figure 1. Herfindahl-Hirschman Index*

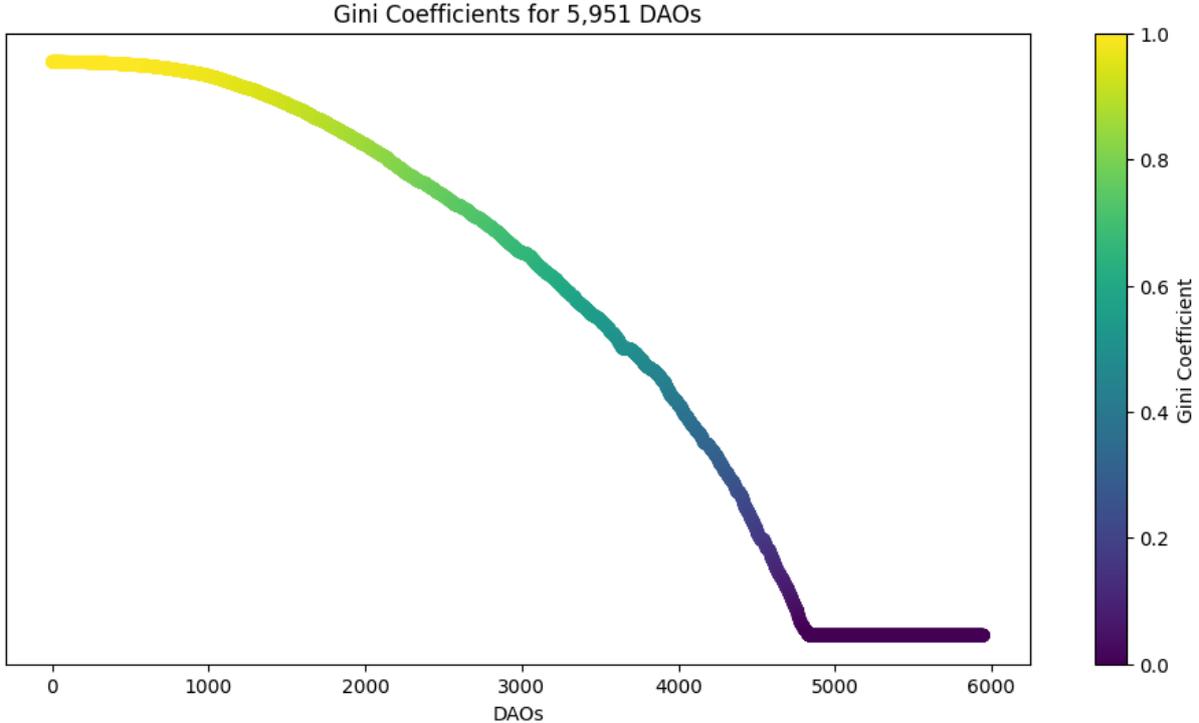

*Figure 2. Gini Coefficients*

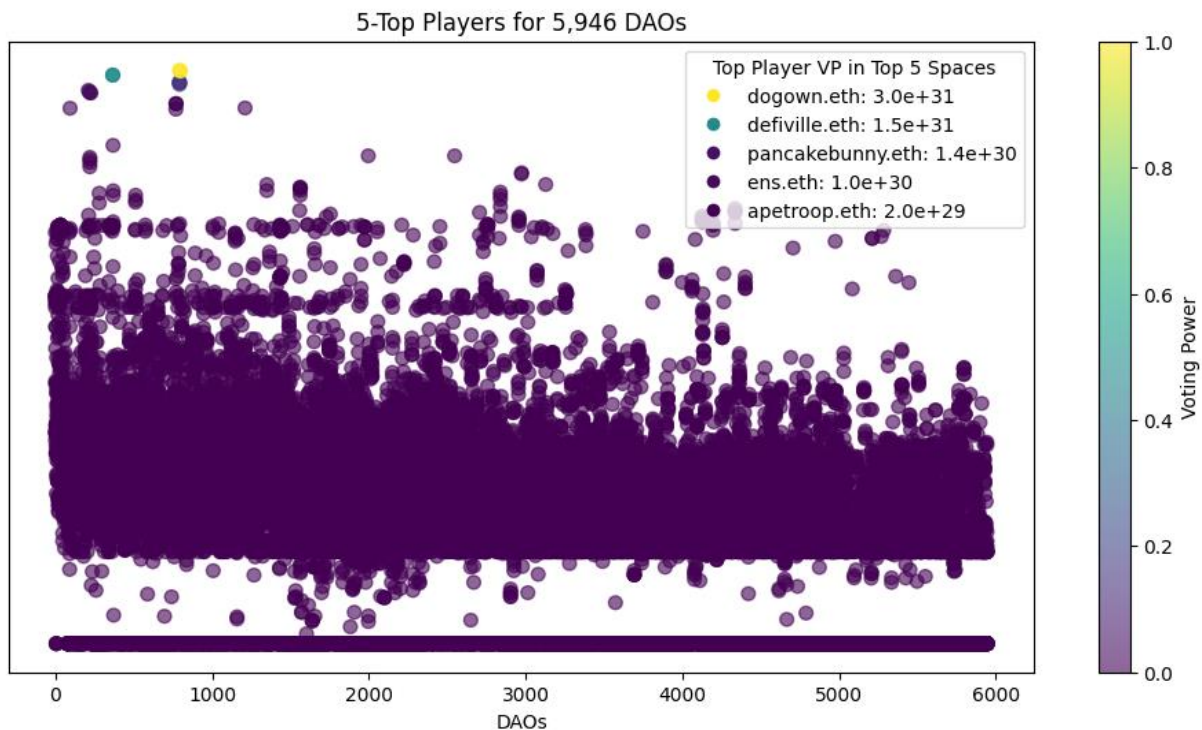

*Figure 3. n-Player Analysis*

Lastly, descriptive statistics on DAOs, such as counts of proposals and votes and comparisons between token prices, may reveal a picture of the current activity of DAOs but they fail to provide a theoretical link to performance, as it has not yet been well demonstrated that many proposals, many voters, or a high token price result in better or worse operational performance for a DAO (the exception in the literature is Rikken et al. (2021), who studied survival rates of DAOs and found that DAOs with more than 20 token holders are more likely to persist).

## 3.3 Working with DAO Data

This section provides a guide for researchers on data collection, selection, and analysis in the context of blockchains and DAOs. It emphasizes the critical distinctions between Mainnet and Testnet blockchains, explores different types of DAOs, and addresses challenges such as token valuation and governance issues. Additionally, the section offers insights into selecting pertinent data, discussing various aspects like the complexities of different DAO models, governance proposal characteristics, levels of community engagement, and the valuation of governance votes in terms of coin voting. The aim is to equip researchers with a foundational understanding of data preparation for effective research in this field.

### 3.3.1 Retrieving Data

Working with quantitative data on DAOs requires proper data cleaning and validation, as with every quantitative analysis. With DAOs, however, there are some less obvious aspects to consider. First, researchers must decide where to get their research data, and a small cottage industry has recently emerged to support these efforts. Prior quantitative research on blockchains was limited by its access to raw blocks (which must be recorded by researchers using custom networking interfaces), lack of standard infrastructure, and the challenges of decoding raw transaction data.

In recent years, several companies have developed tools to simplify these data preparation steps. A prominent source for DAO research is Snapshot,[5] which provides DAO members an integrated interface for DAO voting and offers researchers an API for studying off-chain voting activity by delivering data traces from DAO proposals and votes. DeepDAO[6] provides an analytics engine for researching DAOs, with an estimate of treasury size and voting activity. Dune[7] takes a different approach to analytics by storing all EVM blockchain data in SQL format; this allows researchers to explore chain transactions through traditional SQL queries or by exploring analytical dashboards designed by other researchers in the community. CovalentHQ[8], Moralis[9], and Flipside[10] have similar APIs to gather relevant data on blockchain events, such as on-chain voting, transfers of cryptocurrencies, and token ownership. Coingecko[11] and Coinmarketcap[12] deliver historical and current price data for most cryptocurrencies. Messari[13] has a diverse set of data about token prices, historical token prices, 24-hour volume, token-economics, research insights, and proposals and votes of DAOs. Lastly, data from on-chain DAOs can be collected from The Graph[14], Boardroom[15] or the DAO Analyzer Dataset (Arroyo et al., 2023) which contains insights on aragon[16], daohaus[17], and daostack[18].

Without these APIs, researchers must run blockchain nodes to collect data. For example, a researcher can run an Ethereum Node like Geth[19] to directly retrieve Ethereum blockchain data. However, collecting data directly from the Ethereum blockchain has several downsides, such as the complexity of setting up the node, relatively high hardware requirements, and the need to build an indexer or use an existing one such as eth-indexer[20] to retrieve the required data from the blocks. But, despite the challenges, directly accessing block data has the upside of guaranteeing data accuracy, which is not necessarily true for third party data vendors.

---

[5] https://docs.snapshot.org/tools/api
[6] https://deepdao.io/organizations
[7] https://dune.com/browse/dashboards
[8] https://www.covalenthq.com/
[9] https://moralis.io/
[10] https://docs.flipsidecrypto.com/
[11] https://www.coingecko.com/
[12] https://coinmarketcap.com/
[13] https://messari.io/assets
[14] https://thegraph.com/
[15] https://boardroom.io/
[16] https://aragon.org/
[17] https://daohaus.club/
[18] https://daostack.io/
[19] https://geth.ethereum.org/
[20] https://github.com/getamis/eth-indexer

Supplemental data and auxiliary graphs support deeper analysis. For instance, governance discussions in advance of formal voting are found across the Internet and in hubs like DAOs Discourse.[21] Also, often DAOs and the blockchains they support are developed using principles inherited from several decades of experience with open source software development. This means projects can be analyzed in terms of software engineering by studying version control, typically Github.[22] These changes to technological infrastructure impact social and community changes and through formal change management processes such as Bitcoin Improvement Protocols (BIPs) and Ethereum Improvement Protocols (EIPs), the community guides the development of the DAO platform. These processes mimic earlier Requests for Comments (RFCs) that established the current paradigm of Internet governance.

### 3.3.2 Mainnet and Testnet Data

We must distinguish between Mainnets and Testnets. Mainnets are public blockchains where actual transactions occur, while Testnets are used primarily by developers to test smart contracts, perform token economic simulations, and to upgrade protocols in production-like environments (Kramer, 2023). Testnet DAOs are unsuitable for research as they mostly contain test data. Most DAOs use the Ethereum blockchain. In the Ethereum Virtual Machine (EVM) Environment, every chain has an assigned ChainID[23] to identify it. Examples of Testnet ChainIDs are 3 (Ropsten), 4 (Rinkeby), 5 (Goerli), 97, 280, 595, 1001, 1002, 941, 42, 43113, 69, 5553, 65, 9000, 278611351, and 11297108099. For Mainnets, the Ethereum Mainnet has the ChainID 1, and the Polygon Mainnet has the ChainID 137, for example.

### 3.3.3 Types of DAOs

As Ziegler and Welpe (2022) point out, DAOs follow their own highly individual goals. Categorizing DAOs and understanding their goals remains an ongoing research challenge. DAOs are diverse, as we see in the labels used by the analytics platform Messari[24]: Analytics, Art, City, Culture, DAO Tool, DeFi, Developers, Education, Events/Experiences, Future Of Work, Gaming, Incubator, Infrastructure, Metaverse, Music, NFTs, Pay to Earn (P2E), Public Good Funding, Real World Asset Purchase, Research, Science, Sports, Sustainability and Venture. Yet despite the diversity, common goals and attributes can be identified.

Likewise, the structure of DAOs remains enigmatic. Some may act like traditional organizations with hierarchy and formal responsibilities, others like online communities with weak social ties and open-ended goals, and yet others have criminal and anti-social intents. A key challenge facing the structural categorization of DAOs is the lack of empirically informed organizational theory. In recent years, it has been popular to consider DAOs in terms of stakeholder theory (Freeman, 2010; see Morrison, Mazey, and Wingreen 2020), transaction cost economics (Halaburda et al., 2023; Williamson, 1976), and contract design and agency theory (Milgrom and Roberts, 1992; see M. C. Jensen & Meckling, 2019). Others have drawn from more sociological traditions, considering DAOs from the perspective of Hirschman's (1972) *Exit, Voice, and Loyalty* (see also Schneider & Mannan, 2021) and Ostrom's (1990) *Governing the Commons* (see also DuPont, 2023).

---

[21] https://www.discourse.org/
[22] https://github.com/
[23] https://chainlist.org/
[24] https://messari.io/governor/daos

### 3.4 Difficulties Measuring DAOs

This section explores various challenges to understanding the financial dynamics and governance issues in DAOs. From complications in token pricing to plutocracy, the Sybil problem, and the impact of airdrops and spam proposals, this discussion aims to shed light on the complexities researchers face when measuring DAO governance and financial operations.

#### 3.4.1 Governance Proposals

Despite early efforts to create fully autonomous systems (Pitt et al., 2020), DAO designers quickly realized that change management requires the effective synthesis of humans and technology. And due to the availability of scarce (valuable) tokens associated with DAOs, coin voting on governance proposals quickly became the norm. Naturally, plutocratic governance also quickly emerged, where those with the largest token holdings control the decisions of the DAO. Diverse efforts to curtail or end plutocratic governance include persistent identity systems (Ethereum Name Service, Proof of Participation, Self-Sovereign Identities, and Decentralized Identities) and mechanisms to address Sybil voting (DuPont, 2023). Nonetheless, on-chain and off-chain coin voting on governance proposals remains the central mechanism of change for DAOs and key to our analysis.

However, governance proposals present numerous challenges to researchers. As with all aspects of DAOs, separating the wheat from the chaff is a persistent issue. Aside from the enduring issue of Sybils and their uncontrolled influence on democratic voting, many governance proposals appear to be tests, jokes, or spam. For instance, in a Snapshot.org dataset up to March 28, 2023, we found 3023 proposals that had "test" in their title. Some proposals seem like jokes, but audacity offers context (famously, Constitution DAO attempted to buy an early manuscript of the US Constitution), so whether a DAO really wants to "buy a lambo,"[25] "buy a Space X Rocket,"[26] or "purchase a Lamborghini Revuelto for Marketing Purposes"[27] is an open question. Many proposals are open to voting for short periods (less than 6 hours), suggesting that they are to be interpreted as noise and filtered from analysis (Figure

---

[25] https://snapshot.org/#/dumdave.eth/proposal/0x416d22207321 66296bb6b36af9a02d86f9e9cd247092ab252ceb6a487cdaca05
[26] https://snapshot.org/#/dumdave.eth/proposal/0x8a247337ad9f9f3d54450755b8fdc676372915e6c485da76d005498fd94303a5
[27] https://snapshot.org/#/1inch.eth/proposal/0xabc97c7a6cfa8c8ef0f8024fc4d3f6bbcb186aa47475867df37119d5b520fda8

5Figure 5). On the other hand, Figure 6 reveals extremely long running proposals, which are perhaps evidence of governance proposals being used in unintended or unexpected ways.

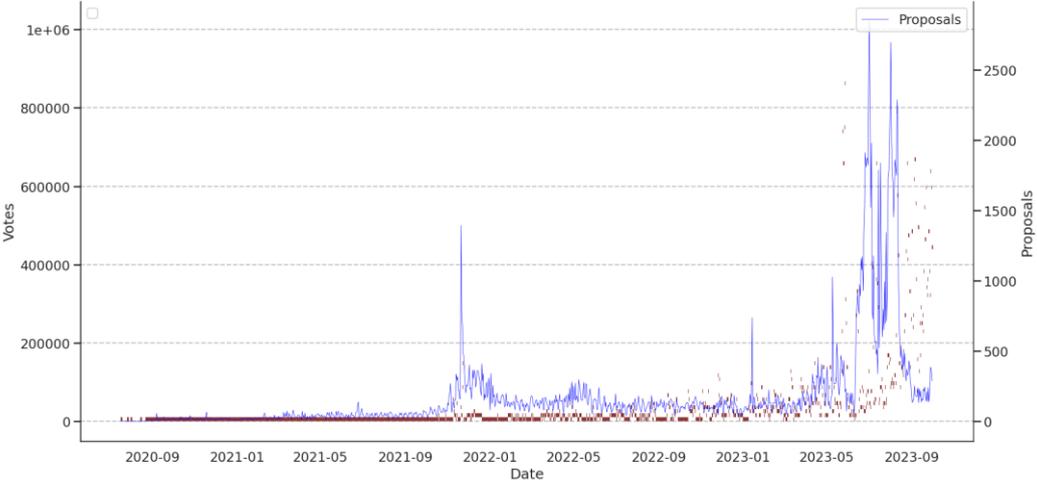

*Figure 4: Votes and proposals created over time*

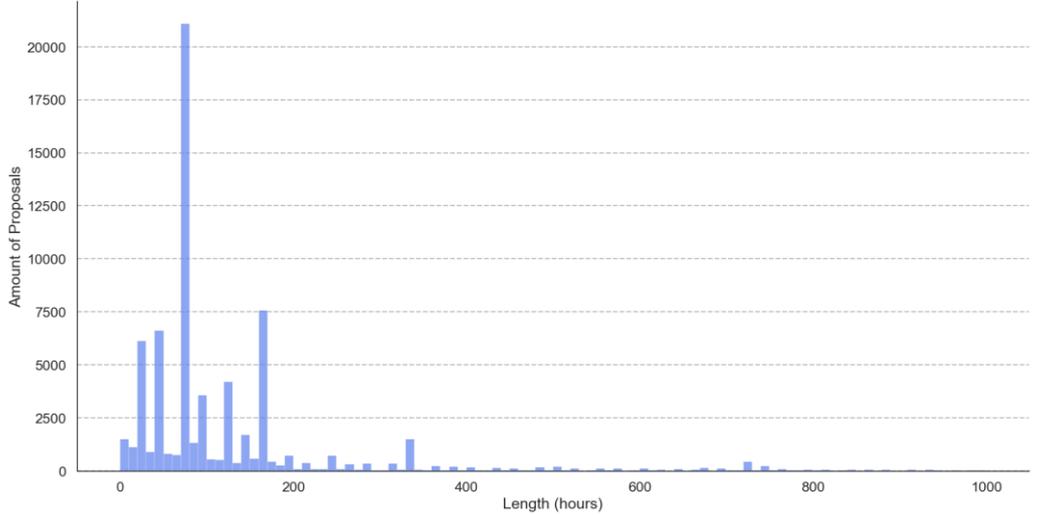

*Figure 5: Filtered proposal durations (short to medium)*

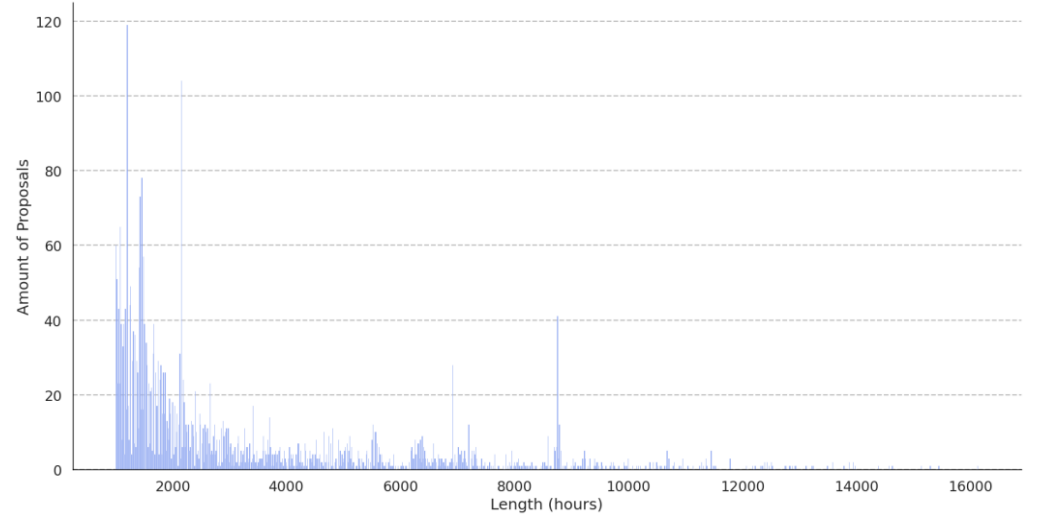

*Figure 6: Filtered proposal durations (long)*

There are many ways to research governance proposals. Proposal discourse may be statistically measured (*n*-grams), topic clustered (supervised and unsupervised methods), analyzed for sentiment (Natural Language Processing), analyzed for importance (Principal Component Analysis), and so on. A newly popular method to discourse analysis uses Large Language Models (LLMs) like ChatGPT to label and cluster proposals. In a small test with LLMs, experimentally we found new categories for DAOs based on proposal discourse: protocol upgrades/expansions, governance and delegation, partnerships and budgeting, risk parameter adjustments, and treasury rebalancing. LLMs can also answer complex questions, such as "does the proposal affect individual wealth?" which opens up many different opportunities for decision support systems and novel research designs.

### 3.4.2 Community Activity

Determining characteristic and exemplary baseline measurements of online community activity is a fraught activity. Far too often researchers implicitly adopt an econometric approach and evaluate DAOs solely by their token prices. Not only are token prices poorly correlated to measurable activities (token prices tend to change in response to bubble dynamics), they also mask many sociologically important phenomena, such as community commitment, motivations, and interactions with other members. For DAOs, a better metric is voting activity, but voting is complicated by test and junk data, Sybils, and the challenges of identifying and tracking anonymous users.

A DAO's activity must be measured against its scale. Very small DAOs with a handful of members and occasional voting will have a distinct structure in comparison to a large DAO with daily or weekly proposals and thousands of participating members. We expect to see strong ties in small networks and weak ties with many absent or invisible ties across large networks (Granovetter, 1973). Moreover, since DAOs are typically permissive in their membership structure (token ownership is usually sufficient to join), individuals may vote across multiple DAOs and take different roles, such as leadership positions. Often, DAOs financially compensate members for contributions, including leadership, which introduces sustainability questions (see Figure 7). As such, many DAOs restrict who can create governance proposals, allowing only known stakeholders or community leaders to create governance proposals. Complicating matters further, many DAOs (like MakerDAO) utilize delegated voting, so key community members may play an outsize role on account of delegated voting power.

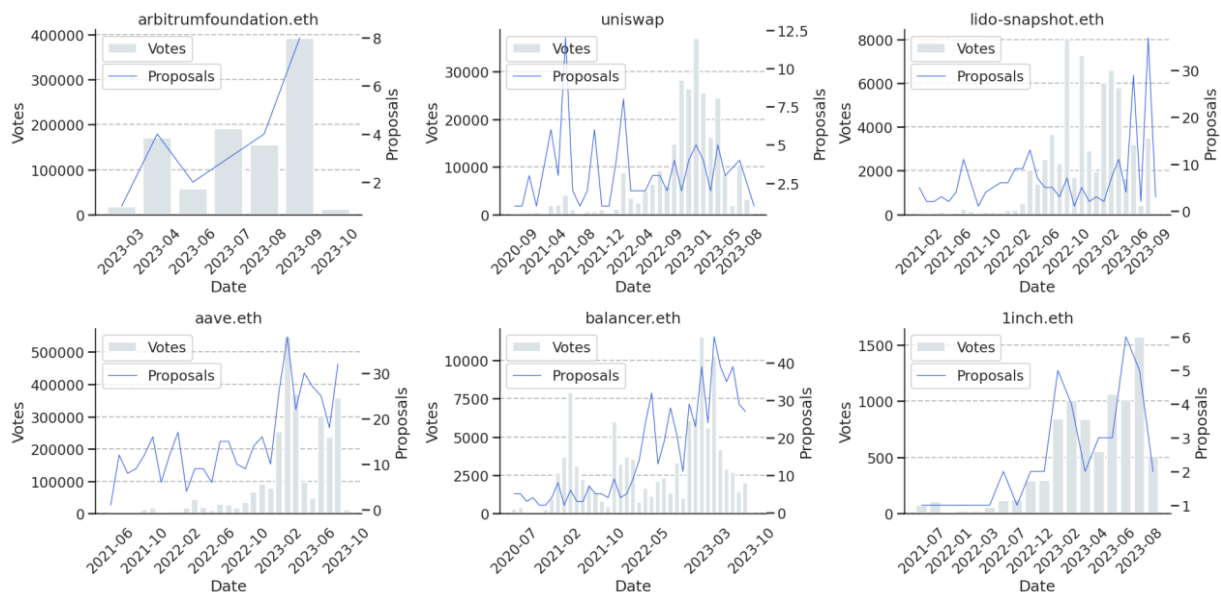

Figure 7: Proposal voting by DAO for major DeFi DAOs

### 3.4.3 Coin Voting

Most DAOs use coin voting. Rather than traditional 'one person, one vote' frameworks, DAOs typically require voters to stake valuable tokens to represent their will. Plutocratic dynamics where large token holders dictate the decisions of the DAO naturally occur, however, the implications of using coin voting in a decentralized environment are not yet well researched. Buterin (2021) highlights issues with decentralized governance (DeGov) in DeFi and offers possible solutions. More broadly, DAOs can be measured and evaluated in terms of treasury size, assets under management (AUM), and total value locked (TVL), although such econometric measurements are necessarily limited, as we discussed above.

### 3.4.4 Token Prices

Not all tokens have monetary value or are tradeable. Non-fungible tokens (NFTs) are singular digital assets that, despite claims about the universal nature of price in neoclassical economics, resist reductionist explanations of their value (see Karpik (2010). Some tokens might even be soulbound, preventing transfers entirely (Weyl et al., 2022). More traditionally, issues with token pricing can be roughly assigned to four classes of error: too little liquidity, untradable or staked tokens, unlisted tokens, and obscure or complex pricing mechanisms.

First, tokens with limited liquidity can experience significant price fluctuations. For tokens with low liquidity, it can be very difficult to discover a price, especially when goods are not homogenous (Milgrom, 2017), which exposes traders to financial risks and introduces analytical complexities. Extreme volatility can render analyses about the effect of token price on proposal outcomes ineffective. More fundamentally, the question of where, how, and why "value" arises is still a conceptual black box.

Second, untradable or staked tokens cannot be used uniformly and users often earmark tokens for specific purposes (Zelizer, 1989). Examples of special purpose tokens include veCRV and sScroll. Unlike traditional monetary instruments, tokens can have mechanisms that decrease or increase each token's voting power (VP) over time, requiring additional data and processing to discover the correct influence of these tokens.

Third, tokens used by new or private DAOs are not yet listed on any central exchange (CEX), such as Coinbase[28] and Binance[29], or any decentralized exchange (DEX), such as Uniswap[30] or SushiSwap[31], making them impossible to price. Fungible tokens may even become non-fungible when engaged in crime; law enforcement in local jurisdictions routinely mark and track illegal transactions and regulatory policies restrict their use.

Fourth, NFTs auctioned on Opensea[32] and similar platforms usually have a floor denominating a minimum price, often corresponding to the price of the last item of the collection sold. However, the next token is not guaranteed to sell for this price, as items are not perfect substitutes. As every NFT is unique, these tokens come with rare attributes that influence the price. Considerable care is required to construct pricing algorithms that are fair and transparent.

### 3.4.5 Token Economics

When conducting financial analyses of DAOs, researchers should examine both on-chain wallets and off-chain bank accounts to accurately assess performance. This dual examination is crucial because DAOs, particularly those operating with both fiat currency and cryptocurrency, can transfer funds between accounts, potentially distorting the financial picture presented by wallet data alone. However, a significant challenge arises in obtaining detailed financial statements from DAOs, including operational costs, income, and salaries. Despite the foundational ethos of transparency and openness in the blockchain and DAO community, the reality is that very few DAOs openly publish their comprehensive business records. This lack of transparency poses a hurdle for researchers seeking to evaluate the true financial health and performance of these organizations.

Additionally, a thorough understanding of token macroeconomic considerations is integral to analyzing DAO finances. This includes examining token burn and mint rates and strategies for token allocations and reserves. To aid in this complex analysis, emerging tools that simulate macroeconomic changes specific to DAOs have become increasingly available. They enable researchers to model various financial scenarios and evaluate the impacts of different economic strategies. Such advancements in simulation tools mark a significant progression in DAO economic research, offering deeper insights into their financial mechanisms and long-term viability. As we discuss below, this analytical capability is essential for navigating the unique and often intricate economic landscapes of DAOs.

---

[28] https://www.coinbase.com/
[29] https://www.binance.com/en
[30] https://uniswap.org/
[31] https://www.sushi.com/
[32] https://opensea.io/

### 3.4.6 The Sybil Problem

The Sybil attack, initially described by Douceur (2002) as "threats from faulty or hostile remote computing elements" on peer-to-peer systems that are "always possible except under extreme and unrealistic assumptions of resource parity and coordination among entities." Due to the anonymity provided by blockchain infrastructure, sybils remain a key problem in real world peer-to-peer system such as Bitcoin and Ethereum. Sybils naturally emerge when new, unique addresses are cryptographically generated from a wallet keychain. Importantly, each new address is cryptographically linked to its generating keychain in ways that only the private key owner can *a priori* determine. And so, transactional privacy is ensured; deanonymizing social networks is infeasible but at its limit deep learning methodologies are chipping away at this anonymity. DuPont (2023) follows recent efforts using deep learning to deanonymize anonymous networks like Bitcoin and Ethereum, using a *k*-anonymity framework (Sweeney, 2002) to identify sybils in DAO governance.

Sybils are particularly problematic for decentralized governance, which relies on coin voting. In a coin voting context, there is no way to ensure democratic processes, which typically require a 'one person, one vote' framework. Instead, due to economic inequality in coin voting, strong plutocratic governance effects emerge. Polycentric governance (DuPont, 2023) offers a flexible community response to sybil attacks, but much more research is necessary to make sybil defense a practical reality for DAO governance.

### 3.4.7 Airdrop Farming and Spam Governance

A unique dimension of Web3 is the presence of so-called Airdrops where specific users are rewarded for early participation, particular actions, or activity in a protocol, blockchain, or decentralized application (dApp) with free tokens that can later be sold for monetary gain or used in governance. Examples of prominent Airdrops are Uniswap, ENS, Bored Ape Yacht Club and Optimism where users have received a non-trivial amounts of tokens (Allen et al., 2023). Out of the twelve notable aidrops that happened between 2014 and 2022, half were retrospective, meaning that the tokens were issued after the project has launched (Allen et al., 2023). Promises of an Airdrop incentivize users to join and participate. For instance, often users must vote on a proposal or otherwise engage with DAO smart contracts to receive a retrospective airdrop. The dispersion of this uncontrolled "free" money often frustrates analysis of DAOs.

# 4 Research Agendas

So far, we have shown how to retrieve, filter, and work with data and identified numerous research gaps, this section offers a perspective on which empirical research questions might be interesting to investigate. Our suggestions are simply a starting point and not exhaustive. We divide the research questions into five themes.

## 4.1 Descriptive Statistics of Online Communities

Descriptive analysis is needed to get an overview of what is currently happening across the DAO landscape. Using time-series data, the evolution of the field over time could be shown. A first glimpse of what this research looks like is provided by Q. Wang et al. (2022), where they descriptively analyze the Snapshot dataset. Since voting occurs in networks, graph analysis also offers many rich opportunities for statistical insight, as (DuPont, 2023) discovered in the Snapshot dataset.

## 4.2 Games, Players, and Institutions

The exploration of games, players, and institutional actors in the context of DAOs, presents a rich area for scholarly inquiry. The impact of prominent venture capital firms like a16z and Paradigm in the DAO landscape is an emerging topic of interest. These firms often have significant token holdings, giving them considerable sway in DAO governance. Their strategies and decision-making processes in this realm are yet to be fully understood by the scientific community. Similarly, the involvement of student clubs such as Penn Blockchain and MIT Blockchain in DAO governance through delegated tokens is a novel development. These groups represent a younger, academically oriented demographic that could have unique motivations and strategies.

Game Theory, as initially proposed by Neumann and Morgenstern (1944), provides a foundational basis for modeling the behavior of these players in DAOs. The application of dynamic games with incomplete information, as conceptualized by Harsanyi (1967), is also particularly relevant. In the context of DAOs, where information asymmetry and changing dynamics are prevalent, these models can help predict and analyze the strategic interactions between different stakeholders.

Moreover, the concept of long-run games, explored by Fudenberg and Levine (2008), is crucial to understanding the sustained interactions and strategies within DAOs. These long-term perspectives can shed light on how continuous participation by known and anonymous entities influences the governance and evolution of DAOs. Their long-term strategies, including investment decisions and voting behaviors, can significantly shape the trajectory of a DAO's development and success.

## 4.3 Models: Correlative and Analytic

In the study of DAOs, the need to develop models is pivotal for advancing our understanding of their dynamics and performance. Correlative models primarily examine statistical relationships between variables. For instance, research leveraging measures such as degree centrality, closeness centrality, eigenvector centrality, and betweenness centrality—concepts foundational to network analysis as discussed in Grassi et al. (2010)—has been instrumental in understanding online communities. However, the application of these measures to DAOs remains underexplored. While these centrality measures are well-established in network theory (see Brede (2012), for an in-depth discussion on network centrality measures), their direct correlation to DAO performance and governance structures is not yet established, leaving a significant gap in the field.

Analytic models, on the other hand, extend beyond identifying correlations to delve into the underlying mechanisms and causal relationships. This approach is crucial in understanding the complexities of DAOs, particularly in aspects like governance and the influence of major stakeholders or 'whales.' While foundational theories in economic governance provide a basis for understanding organizational structures and decision-making processes, their application to the novel context of DAOs remains largely uncharted. The analytic approach could potentially uncover how governance models and the presence of token inequality affect the overall performance and decision-making efficacy within DAOs.

However, both correlative and analytic research in the context of DAOs face a significant challenge due to the absence of a robust theoretical foundation that specifically links these traditional measures and theories to the unique characteristics of DAOs. As a result, there is a pressing need for research that not only applies these established concepts and measures but also adapts and evolves them to suit the distinctive nature of DAOs. This would involve a critical examination of the role of decentralization in performance, the impact of token economics on governance, and the influence of major stakeholders, thereby contributing to a more comprehensive understanding of DAOs in the broader field of blockchain technology and decentralized systems.

### 4.4 Financial Risk and Economic Simulations

Understanding financial risk management in the context of DAO governance is a complex yet crucially understudied topic. Traditional financial risk management tools, both commercial and private, are well-established in the crypto sphere, but their adaptation to DAO governance remains a largely unexplored area. This gap has led companies like Gauntlet to pioneer simulating DAO governance dynamics for risk management. Gauntlet's work focuses on areas such as assessing the impact of adjusting risk parameters in Decentralized Finance (DeFi) DAOs, modelling interactions across different protocols, and optimizing incentive schemes.

Organizations like Token Engineering Commons and Commons Stack have developed tools to support economic and governance simulations. These tools are invaluable to researchers seeking to understand and model the complex dynamics of decentralized governance and their economies. The Augmented Bonding Curve (ABC), developed in collaboration with Block Science, is a form of a bonding curve smart contract that regulates the relationship between token supply and price. It incorporates novel features such as an allowlist with a "trust score" and a common pool treasury, supported by entry and exit tributes. This design aids in establishing a self-sustaining microeconomy within communities, thereby enhancing economic stability and ensuring liquidity in token ecosystems. Complementing ABC, Conviction Voting and Praise emerge as pivotal tools for governance and community engagement. Conviction Voting offers an alternative for collective decision-making by enabling continuous support for proposals, where the conviction or voting power increases the longer tokens are staked. This mechanism democratizes influence, allowing smaller token holders to counterbalance the sway of larger stakeholders, and tackles challenges like governance attacks and voter apathy. On the other hand, Praise serves as a community intelligence system that effectively acknowledges, quantifies, and rewards often overlooked contributions through a peer-to-peer feedback mechanism. This system fosters community engagement and intrinsic motivation among participants.

Additionally, for researchers interested in system dynamics and scenario analysis, cadCAD stands out as a versatile simulation tool. This open-source Python package facilitates the design, testing, and validation of complex systems, integrating with empirical data science practices. With capabilities like Monte Carlo methods, A/B testing, and parameter sweeping, cadCAD is crucial for conducting "what if" analyses, allowing researchers to explore various outcomes and strategies within DAO ecosystems. Together, these tools offer an emerging suite for researchers.

### 4.5 Theory Development, Theory Confirmation, and Paradigm Shifts

Theories that are applicable in other fields, such as online communities in information systems, might also apply to DAOs, such as the Small World Phenomenon (Kleinberg, 2000; Milgram, 1967), Scale-Free Networks (Barabasi & Albert, 1999; Caldarelli, 2008), Community Resilience (Norris et al., 2008), Preferential attachment (Barabasi & Albert, 1999; Kunegis et al., 2013) or information diffusion (Chongfu, 1997). None of these theories have been applied to DAOs yet.

## 5 Conclusion

This agenda highlights the challenges and complexities involved in quantitatively analyzing DAOs. By outlining the limitations of existing metrics and posing critical questions that have yet to be addressed, we provide guidelines for future scholarly research in the evolving field.

The analysis and measurement of DAOs' present a complicated problem. The concept of performance itself is fragmented across different types of DAOs – ranging from investment-centric to community-focused entities. The issue of decentralization adds another layer of complexity; while it is a defining feature of DAOs, its impact on operational performance is still unclear. Even if correctly applied, existing statistical measures may yield insignificant insights due to the diverse nature of DAOs and the problem of retrieving clean and relevant data. As the field matures, datasets used for the research on DAOs should be publicly shared and further refined by future scientists. Furthermore, as the field matures, previous insights may no longer be applicable, as legacy DAOs may no longer be active or may have changed in unforeseen ways.

# 6 References

Allen, D. W., Berg, C., & Lane, A. M. (2023). Why airdrop cryptocurrency tokens? *Journal of Business Research*, *163*, 113945. https://doi.org/10.1016/j.jbusres.2023.113945

Andryukhin, A. A. (2018). Methods of protecting decentralized autonomous organizations from crashes and attacks. *Труды Института Системного Программирования РАН*, *30*(3), 149–164.

Arroyo, J., Davó, D., & Faqir-Rhazoui, Y. (2023). *DAO Analyzer dataset.* https://doi.org/10.5281/zenodo.7669709

Baninemeh, E., Farshidi, S., & Jansen, S. (2023). A decision model for decentralized autonomous organization platform selection: Three industry case studies. *Blockchain: Research and Applications*, *4*(2), 100127. https://doi.org/10.1016/j.bcra.2023.100127

Barabasi, A. L., & Albert, R. (1999). Emergence of scaling in random networks. *Science (New York, N.Y.)*, *286*(5439), 509–512. https://doi.org/10.1126/science.286.5439.509

Beck, R., Müller-Bloch, C., & King, J. L. (2018). Governance in the blockchain economy: A framework and research agenda. *Journal of the Association for Information Systems*, *19*(10), 1.

Bonaventura, M., Ciotti, V., Panzarasa, P., Liverani, S., Lacasa, L., & Latora, V. (2020). Predicting success in the worldwide start-up network. *Scientific Reports*, *10*(1), 345. https://doi.org/10.1038/s41598-019-57209-w

Bose, I., & Pal, R. (2006). Predicting the survival or failure of click-and-mortar corporations: A knowledge discovery approach. *European Journal of Operational Research*, *174*(2), 959–982. https://doi.org/10.1016/j.ejor.2005.05.009

Braun, A., Häusle, N., & Karpischek, S. (2021). Incentivization in Decentralized Autonomous Organizations. *SSRN Electronic Journal.* Advance online publication. https://doi.org/10.2139/ssrn.3760531

Brede, M. (2012). Networks—An Introduction. Mark E. J. Newman. (2010, Oxford University Press.) $65.38, £35.96 (hardcover), 772 pages. ISBN-978-0-19-920665-0. *Artificial Life*, *18*(2), 241–242. https://doi.org/10.1162/ARTL_r_00062

Buterin, V. (2013). *BOOTSTRAPPING A DECENTRALIZED AUTONOMOUS CORPORATION: PART I*. https://bitcoinmagazine.com/technical/bootstrapping-a-decentralized-autonomous-corporation-part-i-1379644274

Buterin, V. (2014). DAOs, DACs, DAs and More: An Incomplete Terminology Guide. https://blog.ethereum.org/2014/05/06/daos-dacs-das-and-more-an-incomplete-terminology-guide

Buterin, V. (2021). *Moving beyond coin voting governance*. https://vitalik.ca/general/2021/08/16/voting3.html

Caldarelli, G. (2008). Scale-free networks: complex webs in nature and technology. *Choice Reviews Online*, *45*(05), 45-2648-45-2648. https://doi.org/10.5860/choice.45-2648

Chao, C.-H., Ting, I.-H., Tseng, Y.-J., Wang, B.-W., Wang, S.-H., Wang, Y.-Q., & Chen, M.-C. (2022). The Study of Decentralized Autonomous Organization (DAO)in Social Network. In *The 9th Multidisciplinary International Social Networks Conference* (pp. 59–65). ACM. https://doi.org/10.1145/3561278.3561293

Chohan, U. W. (2017). The Decentralized Autonomous Organization and Governance Issues. *SSRN Electronic Journal.* Advance online publication. https://doi.org/10.2139/ssrn.3082055

Chongfu, H. (1997). Principle of information diffusion. *Fuzzy Sets and Systems*, *91*(1), 69–90. https://doi.org/10.1016/S0165-0114(96)00257-6

Dhillon, V., Metcalf, D., & Hooper, M. (2017). *Blockchain Enabled Applications*. Apress.

Douceur, J. R. (2002). The Sybil Attack. In G. Goos, J. Hartmanis, J. van Leeuwen, P. Druschel, F. Kaashoek, & A. Rowstron (Eds.), *Lecture Notes in Computer Science. Peer-to-Peer Systems* (Vol. 2429, pp. 251–260). Springer Berlin Heidelberg. https://doi.org/10.1007/3-540-45748-8_24

DuPont, Q. (2017). Experiments in algorithmic governance. In M. Campbell-Verduyn (Ed.), *Bitcoin and Beyond* (pp. 157–177). Routledge. https://doi.org/10.4324/9781315211909-8

DuPont, Q. (2023). *New Online Communities: Graph Deep Learning on Anonymous Voting Networks to Identify Sybils in Polycentric Governance.* https://doi.org/10.48550/arXiv.2311.17929

El Faqir, Y., Arroyo, J., & Hassan, S. (2020). An overview of decentralized autonomous organizations on the blockchain. In G. Robles, K.-J. Stol, & X. Wang (Eds.), *Proceedings of the 16th International Symposium on Open Collaboration* (pp. 1–8). ACM. https://doi.org/10.1145/3412569.3412579

Faqir-Rhazoui, Y., Arroyo, J., & Hassan, S. (2021). A comparative analysis of the platforms for decentralized autonomous organizations in the Ethereum blockchain. *Journal of Internet Services and Applications*, *12*(1). https://doi.org/10.1186/s13174-021-00139-6

Feichtinger, R., Fritsch, R., Vonlanthen, Y., & Wattenhofer, R. (2023). *The Hidden Shortcomings of (D)AOs - An Empirical Study of On-Chain Governance.* https://doi.org/10.48550/arXiv.2302.12125

Freeman, R. E. (2010). *Strategic Management*. Cambridge University Press. https://doi.org/10.1017/CBO9781139192675

Fritsch, R., Müller, M., & Wattenhofer, R. (2022). *Analyzing Voting Power in Decentralized Governance: Who controls DAOs?* https://doi.org/10.48550/arXiv.2204.01176

Fudenberg, D., & Levine, D. K. (2008). *A long-run collaboration on long-run games*. World Scientific.

Goldberg, M., & Schär, F. (2023). Metaverse governance: An empirical analysis of voting within Decentralized Autonomous Organizations. *Journal of Business Research*, *160*, 113764. https://doi.org/10.1016/j.jbusres.2023.113764

Granovetter, M. S. (1973). The Strength of Weak Ties. *American Journal of Sociology*, *78*(6), 1360–1380. https://doi.org/10.1086/225469

Grassi, R., Stefani, S., & Torriero, A. (2010). Centrality in organizational networks. *International Journal of Intelligent Systems*, *25*(3), 253–265. https://doi.org/10.1002/int.20400

Halaburda, H., Levina, N., & Semi, M. (2023). *Digitization of Transaction Terms as a Shift Parameter within TCE: Strong Smart Contract as a New Mode of Transaction Governance*.

Han, J., Lee, J., & Li, T. (2023). DAO Governance. *SSRN Electronic Journal.* Advance online publication. https://doi.org/10.2139/ssrn.4346581

Harsanyi, J. C. (1967). Games with Incomplete Information Played by "Bayesian" Players, I–III Part I. The Basic Model. *Management Science*, *14*(3), 159–182. https://doi.org/10.1287/mnsc.14.3.159

Hassan, S., & Filippi, P. de (2021). Decentralized Autonomous Organization. *Internet Policy Review*, *10*(2). https://doi.org/10.14763/2021.2.1556

Hirschman, A. O. (1972). *Exit, voice, and loyalty: Responses to decline in firms, organizations, and states*. Harvard university press.

Hsieh, Y.-Y., Vergne, J.-P., Anderson, P., Lakhani, K., & Reitzig, M. (2018). Bitcoin and the rise of decentralized autonomous organizations. *Journal of Organization Design*, *7*(1). https://doi.org/10.1186/s41469-018-0038-1

Jensen, J. R., Wachter, V. von, & Ross, O. (2021). *How Decentralized is the Governance of Blockchain-based Finance: Empirical Evidence from four Governance Token Distributions.* https://doi.org/10.48550/arXiv.2102.10096

Jensen, M. C., & Meckling, W. H. (2019). Theory of the firm: Managerial behavior, agency costs and ownership structure. In *Corporate governance* (pp. 77–132). Gower.

Jentzsch, C. (2016). Decentralized autonomous organization to automate governance. *White Paper, November*.

Johnson, R., & Soenen, L. (2003). Indicators of Successful Companies. *European Management Journal*, *21*(3), 364–369. https://doi.org/10.1016/S0263-2373(03)00050-1

Kaal, W. A. (2021). Decentralized Autonomous Organizations: Internal Governance and External Legal Design. *Annals of Corporate Governance*, *5*(4), 237–307. https://doi.org/10.1561/109.00000028

Karpik, L. (2010). *Valuing the Unique*. Princeton University Press. https://doi.org/10.1515/9781400835218

Kleinberg, J. M. (2000). Navigation in a small world. *Nature*, *406*(6798), 845. https://doi.org/10.1038/35022643

Kondova, G., & Barba, R. (2019). Governance of Decentralized Autonomous Organizations. *Journal of Modern Accounting and Auditing*, *15*(8). https://doi.org/10.17265/1548-6583/2019.08.003

Kramer, K. (August 2023). *NETWORKS*. https://ethereum.org/en/developers/docs/networks/#ethereum-mainnet

Kunegis, J., Blattner, M., & Moser, C. (2013). Preferential attachment in online networks. In H. Davis, H. Halpin, A. Pentland, M. Bernstein, L. Adamic, H. Alani, A. Monnin, & R. Rogers (Eds.), *Proceedings of the 5th Annual ACM Web Science Conference* (pp. 205–214). ACM. https://doi.org/10.1145/2464464.2464514

Liu, X. (2023). The Illusion of Democracy? An Empirical Study of DAO Governance and Voting Behavior. *SSRN Electronic Journal.* Advance online publication. https://doi.org/10.2139/ssrn.4441178

Marko, R., & Kostal, K. (2022). Management of Decentralized Autonomous Organizations. In *2022 IEEE International Conference on Omni-layer Intelligent Systems (COINS)* (pp. 1–8). IEEE. https://doi.org/10.1109/COINS54846.2022.9855004

Milgram, S. (1967). The small world problem. *Psychology Today*, *2*(1), 60–67.

Milgrom, P. (2017). *Discovering Prices*. Columbia University Press. https://doi.org/10.7312/milg17598

Neumann, J. von, & Morgenstern, O. (1944). *Theory of games and economic behavior*. *Princeton classic editions*. Princeton Univ. Pr.

Norris, F. H., Stevens, S. P., Pfefferbaum, B., Wyche, K. F., & Pfefferbaum, R. L. (2008). Community resilience as a metaphor, theory, set of capacities, and strategy for disaster readiness. *American Journal of Community Psychology*, *41*(1-2), 127–150. https://doi.org/10.1007/S10464-007-9156-6

Ostrom, E. (1990). *Governing the commons: The evolution of institutions for collective action*. Cambridge University Press.

Pitt, J., Dryzek, J., & Ober, J. (2020). Algorithmic Reflexive Governance for Socio-Techno-Ecological Systems. *IEEE Technology and Society Magazine*, *39*(2), 52–59. https://doi.org/10/gg3qp4

Qin, R., Ding, W., Li, J., Guan, S., Wang, G., Ren, Y., & Qu, Z. (2023). Web3-Based Decentralized Autonomous Organizations and Operations: Architectures, Models, and Mechanisms. *IEEE Transactions on Systems, Man, and Cybernetics: Systems*, *53*(4), 2073–2082. https://doi.org/10.1109/TSMC.2022.3228530

Rikken, O., Janssen, M., & Kwee, Z. (2021). The Ins and Outs of Decentralized Autonomous Organizations (Daos). *SSRN Electronic Journal.* Advance online publication. https://doi.org/10.2139/ssrn.3989559

Santana, C., & Albareda, L. (2022). Blockchain and the emergence of Decentralized Autonomous Organizations (DAOs): An integrative model and research agenda. *Technological Forecasting and Social Change*, *182*, 121806. https://doi.org/10.1016/j.techfore.2022.121806

Schillig, M. (2021). Some Reflections on the Nature of Decentralized (Autonomous) Organizations. *SSRN Electronic Journal.* Advance online publication. https://doi.org/10.2139/ssrn.3915843

Schneider, N., & Mannan, M. (2021). Exit to Community: Strategies for Multi-Stakeholder Ownership in the Platform Economy. *Georgetown Law Technology Review*, *5*(1), 71. https://doi.org/10.31219/osf.io/nmyxp

Sharma, T., Kwon, Y., Pongmala, K., Wang, H., Miller, A., Song, D., & Wang, Y. (2023). *Unpacking How Decentralized Autonomous Organizations (DAOs) Work in Practice*. https://doi.org/10.48550/arXiv.2304.09822

Sun, X., Stasinakis, C., & Sermpinis, G. (2023). *Decentralization illusion in Decentralized Finance: Evidence from tokenized voting in MakerDAO polls*. arXiv.

SWEENEY, L. (2002). k-ANONYMITY: A MODEL FOR PROTECTING PRIVACY. *International Journal of Uncertainty, Fuzziness and Knowledge-Based Systems*, *10*(05), 557–570. https://doi.org/10.1142/S0218488502001648

Torraco, R. J. (2005). Writing Integrative Literature Reviews: Guidelines and Examples. *Human Resource Development Review*, *4*(3), 356–367. https://doi.org/10.1177/1534484305278283

Wang, Q., Yu, G., Sai, Y., Sun, C., Nguyen, L. D., Xu, S., & Chen, S. (2022, November 29). *An Empirical Study on Snapshot DAOs*. http://arxiv.org/pdf/2211.15993v3

Wang, S., Ding, W., Li, J., Yuan, Y., Ouyang, L., & Wang, F.-Y. (2019). Decentralized Autonomous Organizations: Concept, Model, and Applications. *IEEE Transactions on Computational Social Systems*, *6*(5), 870–878. https://doi.org/10.1109/TCSS.2019.2938190

Weyl, E. G., Ohlhaver, P., & Buterin, V. (2022). Decentralized Society: Finding Web3's Soul. *SSRN Electronic Journal.* Advance online publication. https://doi.org/10.2139/ssrn.4105763

Williamson, O. E. (1976). The economics of internal organization: exit and voice in relation to markets and hierarchies. *The American Economic Review*, *66*(2), 369–377.

Wright, A., & Filippi, P. de (2015). Decentralized Blockchain Technology and the Rise of Lex Cryptographia. *SSRN Electronic Journal.* Advance online publication. https://doi.org/10.2139/ssrn.2580664

Wright, A. J. (2021). The rise of decentralized autonomous organizations: Opportunities and challenges.

Zelizer, V. A. (1989). The Social Meaning of Money: "Special Monies". *American Journal of Sociology*, *95*(2), 342–377. https://doi.org/10.1086/229272

Zhao, X., Ai, P., Lai, F., Luo, X., & Benitez, J. (2022). Task management in decentralized autonomous organization. *Journal of Operations Management*, *68*(6-7), 649–674. https://doi.org/10.1002/joom.1179

Ziegler, C., & Welpe, I. M. (2022). A Taxonomy of Decentralized Autonomous Organizations. *ICIS 2022 Proceedings*. https://aisel.aisnet.org/icis2022/blockchain/blockchain/1

Ziegler, C., & Zehra, S. R. (2023). Decoding Decentralized Autonomous Organizations: A Content Analysis Approach to Understanding Scoring Platforms. *Journal of Risk and Financial Management*, *16*(7), 330. https://doi.org/10.3390/jrfm16070330